\documentclass[aps,pra,twocolumn,showpacs,amsmath,amssymb,superscriptaddress,superscriptaddress]{revtex4-1}

\usepackage{graphicx}
\usepackage{verbatim}
\usepackage{epstopdf}
\usepackage{bm}
\usepackage{times}
\usepackage{comment}
\usepackage{dsfont}
\usepackage{color}
\definecolor{blue}{RGB}{0,0,255}
\usepackage[colorlinks,citecolor=blue,linkcolor=blue,urlcolor=blue]{hyperref}
\pdfoutput=1

\begin{document}

\def\BE{\begin{equation}}
\def\EE{\end{equation}}
\def\BY{\begin{eqnarray}}
\def\EY{\end{eqnarray}}
\def\BI{\begin{itemize}}
\def\EI{\end{itemize}}
\def\L{\label}
\def\nn{\nonumber}
\def\({\left (}
\def\){\right)}
\def\[{\left [}
\def\]{\right]}
\def\<{\langle}
\def\>{\rangle}
\def\o{\overline}
\def\BA{\begin{array}}
\def\EA{\end{array}}
\def\dsp{\displaystyle}
\def\ds{\displaystyle}
\def\k{\kappa}
\def\dd{\delta}
\def\D{\Delta}
\def\w{\omega}
\def\W{\Omega}
\def\v{\nu}
\def\a{\alpha}
\def\b{\beta}
\def\e{\varepsilon}
\def\ee{\text{e}}
\def\d{\partial}
\def\g{\gamma}
\def\G{\Gamma}
\def\tt{\theta}
\def\t{\tau}
\def\+{\dag}
\def\8{\infty}
\def\x{\xi}
\def\l{\lambda}
\def\ii{\text{i}}
\def\={\approx}
\def\xc{\frac{2x}{c}}
\def\->{\rightarrow}
\def\r{\vec{r}}
\def\k{\vec{k}}
\def\sinc{\mathrm{sinc}}
\def\xx{\textbf{x}}
\def\yy{\textbf{y}}
\def\qq{\textbf{q}}
\def\rr{\boldsymbol{\rho}}
\newcommand{\ud}{\,\mathrm{d}} 
\def\out{|\textrm{out}\rangle}
\def\inn{|\textrm{in}\rangle}
\def\sqz{|\textrm{sqz}\rangle}
\def\A{\text{A}}
\def\B{\text{B}}
\def\AB{\text{AB}}
\def\where{\text{where:}}
\def\F{{\cal F}}
\def\A{{\cal A}}
\def\si{\text{si}}
\def\s{\text{s}}
\def\i{\text{i}}
\def\c{\text{c}}
\def\u{\text{u}}
\def\m{\text{m}}
\def\X{\textcolor{red}{XXX}}
\title{Efficient generation of temporally shaped photons using nonlocal spectral filtering}

\author{Valentin Averchenko}
\email{valentin.averchenko@gmail.com}
\affiliation{St. Petersburg State University, Ul'yanovskaya str. 3, 198504 Sankt-Peterburg, Russia}

\author{Denis Sych}
\affiliation{P. N. Lebedev Physical Institute, Russian Academy of Sciences, Leninskiy Prospekt 53, 119991 Moscow, Russia}
\affiliation{Russian Quantum Center, Skolkovo, 143025 Moscow, Russia}
\affiliation{Moscow Pedagogical State University, 29 Malaya Pirogovskaya St, 119435 Moscow, Russia}

\author{Christoph Marquardt}
\affiliation{Max-Planck-Institut for the Science of Light, Staudtstra{\ss}e 2, 91058 Erlangen, Germany}
\affiliation{Institute for Optics, Information and Photonics, University Erlangen-N\"{u}rnberg, Staudtstra{\ss}e 7/B2, 91058 Erlangen, Germany}

\author{Gerd Leuchs}
\affiliation{Max-Planck-Institut for the Science of Light, Staudtstra{\ss}e 2, 91058 Erlangen, Germany}
\affiliation{Institute for Optics, Information and Photonics, University Erlangen-N\"{u}rnberg, Staudtstra{\ss}e 7/B2, 91058 Erlangen, Germany}

\date{\today}

\begin{abstract}

We study the generation of single-photon pulses with the tailored temporal shape via nonlocal spectral filtering.
A shaped photon is heralded from a time-energy entangled photon pair upon spectral filtering and time-resolved detection of its entangled counterpart.
We show that the temporal shape of the heralded photon is defined by the time-inverted impulse response of the spectral filter and does not depend on the heralding instant.
Thus one can avoid post-selection of particular heralding instants and achieve substantially higher heralding rate of shaped photons as compared to the generation of photons via nonlocal temporal modulation.
Furthermore, the method can be used to generate shaped photons with a coherence time in the ns-$\mu$s range and is particularly suitable to produce photons with the exponentially rising temporal shape required for efficient interfacing to a single quantum emitter in free space.

\end{abstract}


\maketitle

\section{Introduction}

Single-photon optical pulses with the desired temporal shape, i.e. desired temporal envelope of the electro-magnetic field, can be produced from time-energy entangled photon pairs either via nonlocal temporal \cite{Averchenko2017} or spectral \cite{Klyshko1998,Bellini2003, Gulati2014, Du2015, Sych2017} modulation 
\footnote{Here we consider only single-photon shaping methods which use entangled photon pairs. A short overview of other methods can be found in \cite{Averchenko2017}.}.
Both shaping methods can be implemented using various sources of entangled photon pairs and can be potentially scalable to produce multiple shaped photons.
Also, both methods are seemingly highly related to each other and can be described in a unified formalism \cite{Sych2017}. One might even think that it does not matter which modulation one uses on a given entangled resource, except for the difference in the shaping procedure.
But there is an important difference between these methods on which we focus in this work.

\begin{figure*}	\center{\includegraphics[width=0.7\linewidth]{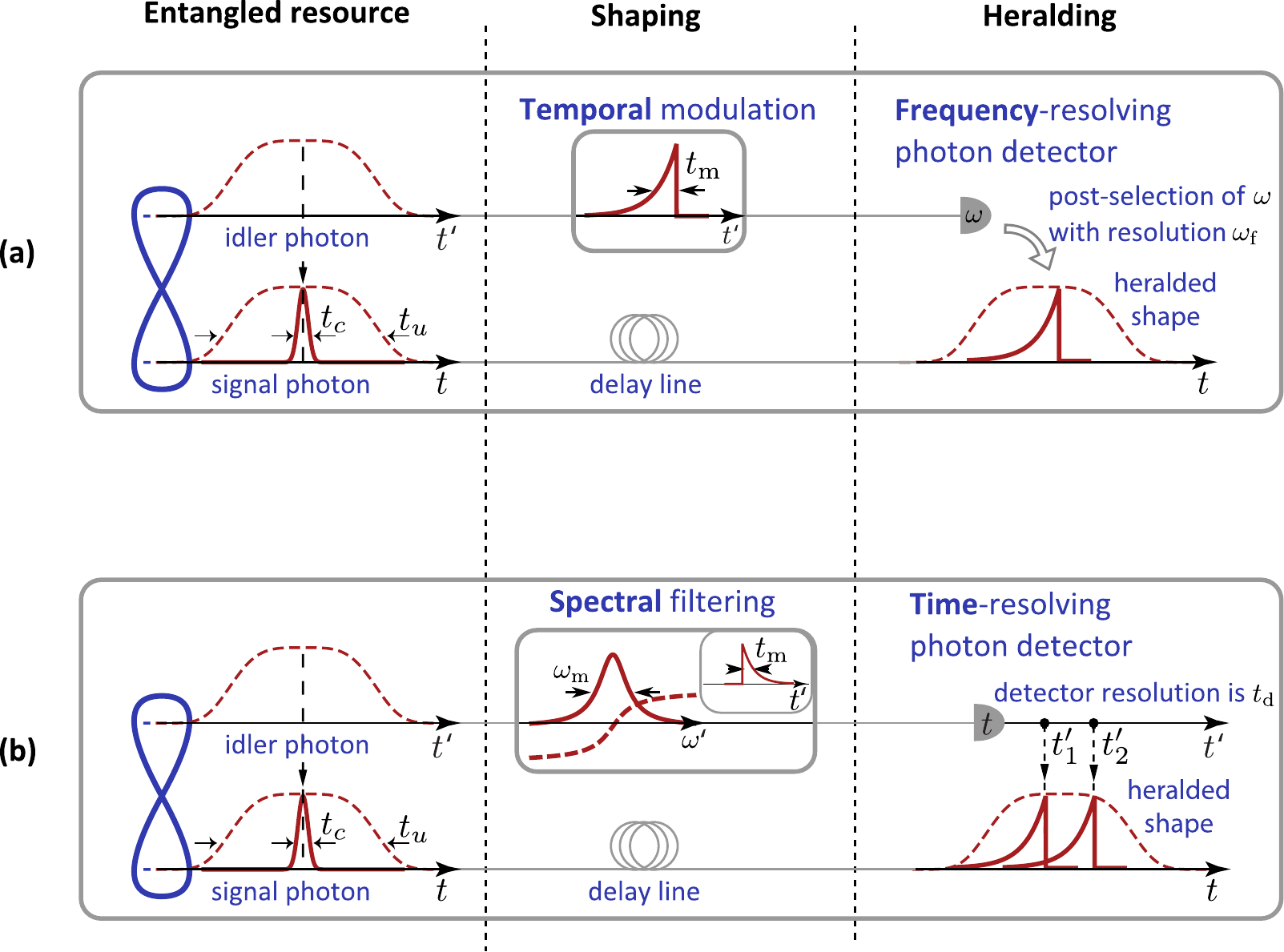}}
	\caption{
Generation of temporally shaped single-photon pulses from time-energy entangled photon pairs via (a) nonlocal temporal modulation or (b) nonlocal spectral filtering. As an example, the generation of photons with the exponentially rising temporal shape is considered. Time-energy entangled photons are called signal and idler, for convenience. 
The photons are correlated in the time domain, i.e. their conditional temporal width $t_\c$ is smaller than unconditional width $t_\u$: $t_\c \ll t_\u$.
In (a) idler photon is modulated with the exponentially rising wave-form, with the characteristic modulation time $t_\m$.
In (b) idler photon passes spectral filter with the Lorentzian transmission function with the characteristic bandwidth $\w_\m \sim t_\m^{-1}$.
Inset shows the corresponding amplitude (solid line) and phase (dashed line) of the impulse response of the spectral filter.
In the first case, post-selection of the particular outcome of the frequency-resolving detector is required to herald a signal photon with the desired temporal shape and carrier frequency. The post-selection is performed with a spectral resolution $\w_\text{d}$.
In the second case, detection instant defines timing of the heralded shaped photon. 
Characteristic temporal resolution of the detector is $t_\text{d}$.
All clicks herald signal photons with identical temporal shape and can be used without post-selection.
}
	\label{fig:scheme}
	\end{figure*}

When using the first method (see Fig.~\ref{fig:scheme}(a)) the shaped photon is produced by temporal modulation and frequency resolving detection of its entangled counterpart.
The modulation function uniquely defines the temporal shape of the heralded photon, while the outcome of the frequency-resolving detector defines the central frequency of the heralded photon \cite{Averchenko2017}.
Therefore, post-selection on the detected frequency is required to produce a shaped photon with the desired central frequency, which is important for the efficient interaction of the photon with a single two-level atom \cite{Stobinska2009, Wang2011, Leong2016}, or a quantum memory \cite{Novikova2007, Farrera2016}.
However, such post-selection lowers the generation rate of shaped photons and limits the applicability of this shaping method.

We show in this work that the situation is different when the shaped photon is produced from an entangled photon pair via spectral filtering and time-resolved detection of its entangled counterpart  (see Fig.~\ref{fig:scheme}(b)).
The filter function uniquely defines the spectral shape of the photon and its central frequency, and the heralding instant defines the timing of the produced shaped photon.
Heralding instants are not known in advance, but they can be used as synchronization signals for further applications of heralded shaped photons.
For example, when a heralded photon is supposed to be stored in a quantum memory, then the storage process can be synchronized using a heralding instant of the photon. 
To be more specific, the heralding instant of the shaped photon can be used  as a synchronization signal in the case of a quantum memory based on a three-level scheme \cite{Novikova2007}, where the control pulse needs to be properly  synchronized with a single-photon pulse.
Therefore, all heralding events can be used without the need of post-selection.
Compared to the case of temporal modulation the generation rate of shaped photons can be substantially increased.

In this paper we theoretically analyze the generation of shaped photons from time-energy entangled photon pairs using a nonlocal spectral filtering method. 
We demonstrate its main features and, particularly, estimate the heralding rate of shaped photons achieved with the method.
In Sec.~\ref{simple_model} we present a simple description of the method assuming perfect temporal correlations of photon pairs.
In Sec.~\ref{sec:arb} we derive a general expression for the shape of the photon heralded from photon pairs with the finite correlation time and estimate the heralding probability of photons. 
In Secs.~\ref{sec:exponent} and \ref{cw} we model the generation of photons with an exponentially rising temporal shape from practical source of entangled photon pairs, such as cavity-assisted spontaneous nonlinear process.
In Sec.~\ref{sec:atom} we consider an interaction of a single photon and a two-level atom and estimate maximum excitation probability of the atom, when a photon has an exponentially rising temporal shape.
In Sec.~\ref{sec:imperfections} we consider the effect of potential experimental imperfections on the shaping performance and in Sec.~\ref{sec:estimations} we estimate the application range of the shaping method.
Sec.~\ref{sec:conclusion} concludes the paper.

\section{Shaping with maximally entangled photons}\L{simple_model}

Here we present a simple model of the generation of temporally shaped photons from entangled photon pairs using nonlocal spectral filtering
\footnote{The analysis of the photon shaping via the nonlocal temporal modulation shown in Fig.~\ref{fig:scheme}(a) has been performed in \cite{Averchenko2017}.}.
Principal scheme of the shaping procedure is shown in Fig.~\ref{fig:scheme}(b).
We call photons of an entangled pair as signal and idler and approximate the entangled state of a photon pair in the time domain as follows:
	\begin{align}
	& |\Psi\>_\si \propto \int \ud t |t\>_\s |t\>_\i, 
	\L{tt}
	\end{align}
where $|t\> = {\hat a}^\+(t)|0\>$ stands for a single-photon state.
An optical pulse in such a quantum state causes the detector to trigger at the moment of time $t$.
Hereafter, integration limits are infinite.

The expression (\ref{tt}) implies that the detection moment of signal/idler photon is uncertain, but these moments are perfectly correlated.
Making the substitution $|t\> \propto \int \ud \w \; \ee^{\ii \w t} |\w\>$ for the signal and idler single-photon states in (\ref{tt}), one gets the following representation of the entangled state in the frequency domain:
	\begin{align}
	& |\Psi\>_\si \propto\int \ud \w |\w\>_\s |-\w\>_\i,
	\L{ww}
	\end{align}
where $|\w\> = {\hat a}^\+(\w)|0\>$ denotes a single-photon state with the photon frequency $\w$. 
Hereafter, frequencies of signal and idler fields are counted relative to corresponding central frequencies of fields.
The expression (\ref{ww}) shows that entangled photons which are perfectly correlated in time also possess perfect anti-correlations in frequency.   
The physical reason is the following - in a physical process of the photon pair generation the photons are generated instantly (hence perfect correlation in time) and their total energy is well-defined (hence perfect anto-correlation in frequency).

At the next step of the shaping method the idler photon is sent to the spectral filter. The effect of the filter on the photon is described in the frequency domain by a complex transmission function $F(\w)$. It gives a probability amplitude for a photon with a frequency $\w$ to pass the filter.
In the time domain the filter is described by the impulse response function $\F(\t) = \int F(\w) \ee^{-\i \w\t} \ud\w/2\pi$, which defines the probability amplitude for a single photon $|\t\>$ to pass the filter with the time delay $\t$. If the idler photon passes the filter, then the resulting quantum state of the photon pair is obtained from (\ref{tt}) and (\ref{ww}) using the following substitutions:
	\begin{align}
	& |\w\>_\i \rightarrow F(\w) |\w\>_\i,\\
	& |t\>_\i \rightarrow \int \ud\t \F(\t) |t+\t\>_\i \L{t_filt}
	\end{align}
A photocount of the detector at the time instant $t'$ corresponds to the detection of an idler photon in the state: $|t'\>_\i \propto \int \ud\w' \ee^{\ii\w't'} |\w'\>_\i$.
This detection event heralds a signal photon in the following conditional quantum state:
	\begin{align}
	& |\psi\>_\s \propto {}_\i\<t'|\Psi\>_\si \propto \int\ud\w F(-\w)\ee^{\ii\w t'}|\w\>_\s.
	\L{shape_w}
	\end{align}
The last expression shows that the spectral shape of the heralded photon is defined by the frequency-reversed filter function $F(\w)$. 
The inversion is due to the initial frequency anti-correlations  of signal and idler photons (\ref{ww}).
Applying the Fourier transformation to (\ref{shape_w}), one gets the corresponding conditional state of a signal photon in the time domain
	\begin{align}
	& |\psi\>_\s \propto \int\ud t \F(t'-t)|t\>_\s.
	\L{shape_t}
	\end{align}
The expression shows that the temporal shape of the heralded photon is defined by time-reversed impulse response of the filter: $\psi_s(t|t') \propto \F(t'-t)$.
The following distinctive features of the shaping method are related to this result.

First, the temporal shape of the heralded signal photon can be controlled by choosing the spectral filter with a specific spectral function.
The spectral filter can be implemented with an optical interferometer \cite{Bellini2003, Zavatta2006}, optical cavities
\cite{Haase2009, Gulati2014, Qin2015, Monken1993}, 
an atomic ensemble \cite{Wolfgramm2011, Schunk2015}, etc.

Second, the heralding instant $t'$ defines the timing of the heralded shaped photon.
This is illustrated in Fig.~\ref{fig:scheme}(b), where detector photocounts at different time instants herald photons which have identical shapes and are synchronized with the heralding instants. 
Therefore, post-selection of particular heralding instants is not required, and a higher heralding rate of shaped photons can be achieved.
This property of shaping via the spectral filtering makes it substantially more practical compared to the photon shaping via the temporal modulation \cite{Averchenko2017}. 

The third feature is the time reversal of the filter response which can be useful for several quantum optics tasks. For example, one can generate a single photon with the rising exponential profile using a spectral filter with the exponentially decaying impulse response which can be implemented with the combination of optical cavities \cite{Haase2009}.
Such shaped photons are required for the efficient excitation of a two-level quantum emitter in free space \cite{Stobinska2009, Wang2011, Leong2016}, photon coupling to an atomic ensemble \cite{ZhangPRL2012}, and efficient photon loading into an optical cavity \cite{Bader2013, Liu2014}.

\section{Shaping with non maximally entangled photons. Heralding probability}\L{sec:arb}

	\begin{figure*}
	\center{\includegraphics[width=0.8\linewidth]{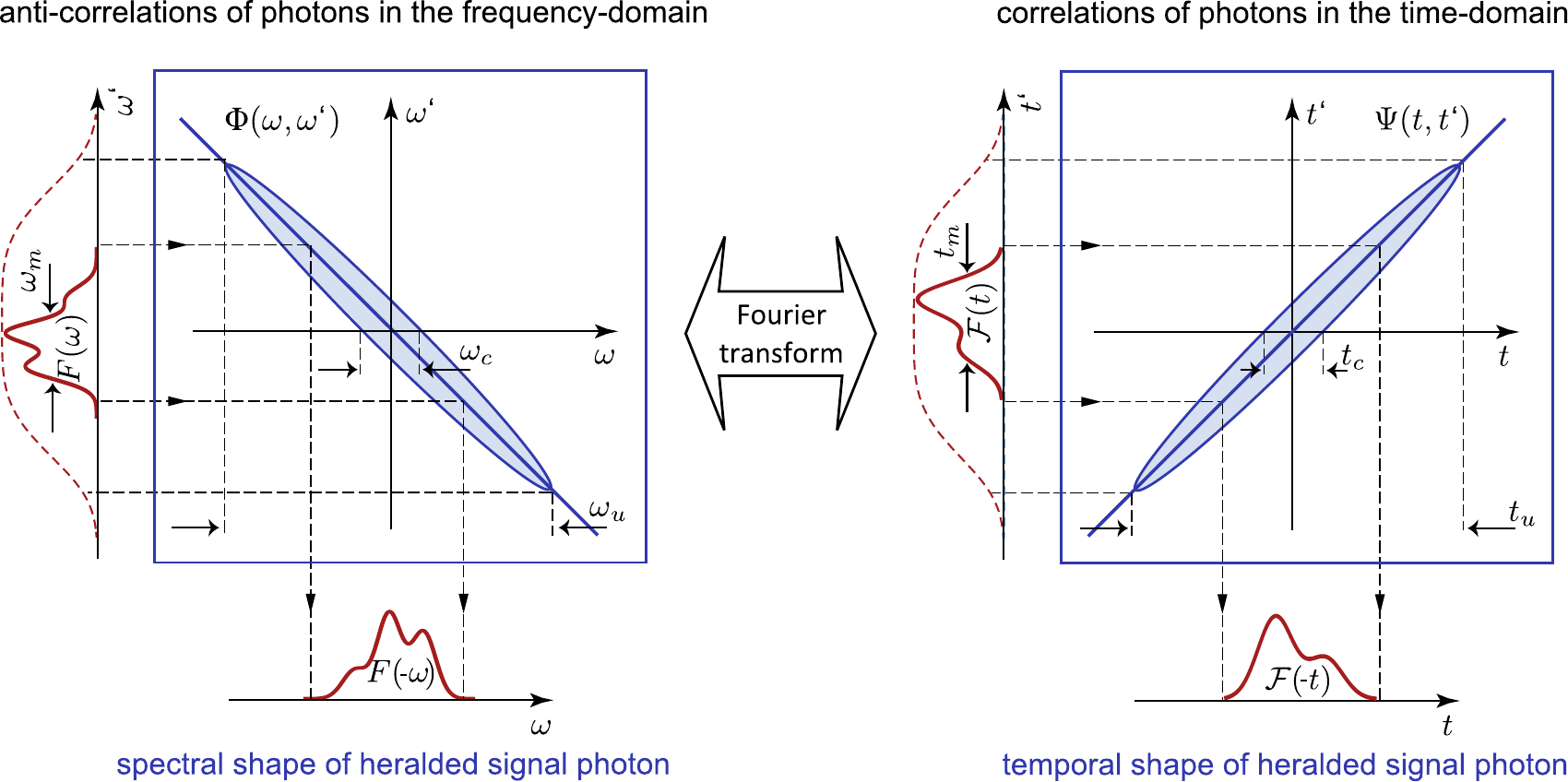}}
	\caption{
Joint probability densities of entangled photons are represented by blue ellipses in frequency (left) and time (right) domains.
$\w_\u, t_\u$ denote unconditional spectral and temporal width of signal/idler photons and $\w_\c, t_\c$ denote conditional ones. 
The ellipses tend to anti-diagonal/diagonal lines in the limit of perfectly correlated photons. 
Spectral filtering of the idler photon and its successful time-resolved detection heralds shaped signal photon such that: its spectral shape is defined by the frequency-reversed transmission function of the filter (see (\ref{shape_w})) and its temporal shape is defined by the time-reversed impulse response of the filter (see (\ref{shape_t})).}
	\label{fig:shape}
	\end{figure*}

Above we analyzed the generation of shaped signal photons assuming that the entangled photons are perfectly correlated in time and anti-correlated in frequency (see (\ref{tt}) and (\ref{ww})).
Here we consider a more realistic case, when the correlations of photons are non-perfect,
and we derive a more general expression for the shape of signal photons produced via nonlocal spectral filtering.
We will also derive the conditions necessary for the method to work and we estimate the heralding probability of shaped photons.

The general time-energy entangled state of two photons reads as
\footnote{For example, see \cite{Law2000} for the description of time-energy entangled photons generated in a process of parametric down-conversion.}:
	\begin{align}
	|\Psi\>_\si &= \iint \ud t\ud t' \Psi(t,t') |t\>_\s |t'\>_\i \\
	&= \iint \ud \w\ud \w' \Phi(\w,\w') |\w\>_\s |\w'\>_\i.
	\end{align}
Here $\Psi(t,t')$ and $\Phi(\w,\w')$ are temporal and spectral probability amplitudes. Their squared modulus defines the joint probability density to detect signal and idler photons at time instants $t$ and $t'$ or with frequencies $\w$ and $\w'$, respectively. The amplitudes are related via the two-dimensional Fourier transformation.
Fig.~\ref{fig:shape} shows schematically the joint probability densities in the time and frequency domains. 
The functions are stretched along diagonals, that means the photons are correlated in time and anti-correlated in frequency.
$t_\text{u,c}$ and $\w_\text{u,c}$ depict characteristic unconditional/conditional temporal and spectral widths of photons.
These widths are related via the Fourier transformation and satisfy the following uncertainty relations \cite{AliKhan2006, Howell2004}: $t_\text{u} \w_\text{c} \sim 1$ and $t_\text{c} \w_\text{u} \sim 1$.
Thus, correlations of photons in time ($t_\u \gg t_\c$) imply their anti-correlations in frequency ($\w_\u \gg \w_\c$).
Also $\w_\m$ and $t_\m$ denote the characteristic passband of the filter and its corresponding response time. These quantities fulfill the uncertainty relation $t_\m \w_\m \sim 1$.

Filtering of the idler photon is described by the following transformation of spectral and temporal joint probability amplitudes, respectively
	\begin{align}
	& \Phi(\w,\w') \rightarrow F(\w')\Phi(\w,\w'), 
	\L{Phi_ww}
	\\
	& \Psi(t,t') \rightarrow \int\ud \t \F(\t) \Psi(t,t'-\t).
	\L{Psi_tt}
	\end{align}
The detection of the idler photon at time instant $t'$ announces the signal photon in the conditional state. The state reads in the frequency and temporal domains, respectively
	\begin{align}
	|\psi\>_\s &\propto {}_\i\<t'|\Psi\>_\si \nonumber\\
	 &\propto\int\ud\w \left[\int\ud\w' F(\w') \Phi(\w,\w') \ee^{-\ii \w' t'}\right] |\w\>_\s
	\L{psi_w-1}
	\\
	& = \int\ud t \left[\int\ud\t\F(\t) \Psi(t,t'-\t)\right] |t\>_\s.
	\L{psi_t-1}
	\end{align}
The last two expressions are generalizations of the expressions (\ref{shape_w}) and (\ref{shape_t}), respectively, for the case of arbitrarily entangled photon pairs.
The expressions in square brackets give the shape of the heralded signal photon, in the frequency and time domains, respectively.
In general, the shape depends on the initial entanglement of photons, filter function and heralding instant.

In the limit of perfectly correlated photons, the correlation time $t_\text{c}$ of photons is the smallest parameter in a shaping experiment and unconditional temporal spread $t_\text{u}$ is the largest one, i.e.:
	\begin{align}
	& t_\text{c} \ll t_\text{m} \ll t_\text{u} \L{t-cond},
	\end{align}
where $t_\text{m}$ is the characteristic response time of the filter. 
The equivalent inequality in the frequency domain is: 
	\begin{align}
	& \w_\text{c} \ll \w_\text{m} \ll \w_\text{u}.
	\end{align}
It means that the filter pass band $\w_\m$ is larger than conditional spectral spread of photons $\w_\c$ and smaller then the unconditional one $\w_\u$. 
Corresponding probability densities tend to blue straight diagonal lines in Fig.~\ref{fig:shape}.
Using the approximations  $\Psi(t,t') \propto \dd(t-t')$ and $\Phi(\w,\w') \propto \dd(\w+\w')$ in (\ref{psi_w-1}) and (\ref{psi_t-1}) one reproduces the results of the previous section, namely, the photon shape is defined by the filter function (see Eqs. (\ref{shape_w}) and (\ref{shape_t})).

Assume that after the filter all idler photons are used as heralds for shaped signal photons, i.e. no post-selection of detection outcomes is performed (see Fig.~\ref{fig:scheme}(b)).
Then the heralding probability of shaped photons is defined by the fraction of idler photons that have passed the filter. It can be formally calculated as
	\begin{align}
	R &= \iint\ud\w\ud\w' \left|F(\w') \Phi(\w,\w')\right|^2 \L{RR}\\
	&\approx \frac{\w_\m}{\w_\u} = \frac{t_\c}{t_\m}. \L{wm/wu}
	\end{align}
The last two expressions are estimations of the heralding probability using characteristic spectral and temporal widths. Heralding probability is a probability that photons with unconditional spectral
width $\w_\text{u}$ will pass a filter with the passband $\w_\text{m}$.

For the comparison, the heralding probability of shaped photons produced via the temporal modulation (see Fig.~(\ref{fig:scheme})(a) and \cite{Averchenko2017}), can be estimated as: $R' \approx \frac{t_\m}{t_\u} \w_\text{f} t_\c$.
Here, the ratio $t_\m/t_\u \leqslant 1$ is a probability that idler photon with unconditional temporal spread $t_\u$ passes the temporal modulator within the time window $t_\m$.
Second term $\w_\text{f} t_\c \ll 1$ represents the reduction of heralding probability due to the selection of outcomes of the frequency-resolving detector within the narrow frequency range $\w_\text{f} \ll t_\m$. Such post-selection is required to produce photons with a well-defined central frequency.
Since this term is absent in expression (\ref{wm/wu}) the heralding probability of the shaped photon achieved via nonlocal spectral modulation can be substantially higher than the heralding probability provided by nonlocal temporal modulation method.
Namely, the enhancement factor is $ R/R' \approx 1/\w_\text{f} t_\m \gg 1$.
Below we consider the generation of shaped photons using a particular source of entangled photons.

\section{Generation of photons with an exponentially rising temporal shape from single photon pairs}\L{sec:exponent}

Here we consider the generation of photons with the exponentially rising temporal shape, which are required for the efficient interaction with a two-level quantum emitter in free space \cite{Stobinska2009, Wang2011, Leong2016}, with an ensemble of atoms \cite{ZhangPRL2012}, for coupling to a quantum memory \cite{Novikova2007, Farrera2016}, to an optical \cite{Bader2013} or a superconducting resonator \cite{Wenner2014}.
As a resource, we consider pairs of time-energy entangled photons with the following joint probability amplitude of a single pair:
	\begin{align}
	\begin{split}
	\Psi(t,t') \propto \ee^{-|t-t'|/2 t_\text{c}},\\
	\text{where}\; 0 \leq t,t' \leq t_\text{u}
	\end{split}
	\L{psi_spdc}
	\end{align}
This joint amplitude shows that the probability to detect a signal/idler photon is non-zero and constant within the time interval of duration $t_\u$. 
Signal and idler detection instants are correlated with the correlation time $t_\c$.
Such entangled photon pairs can be produced in cavity-assisted spontaneous parametric down-conversion (SPDC) \cite{Fortsch2013}, or spontaneous four-wave mixing (SFWM) \cite{Lu2016}.
For these sources $t_\c$ is defined by the photon lifetime in the cavity.
$t_\u$ depends on the generation process. It is defined by the duration of the pump pulses that drive the nonlinear process in the pulsed pump regime.
In the continuous pump regime $t_\u$ is defined by the average time interval between successively generated photon pairs and can be estimated as the inverse average rate of photon pairs $\bar n^{-1}$.

To produce a signal photon of exponentially rising temporal shape with the rise time $t_\text{m}$, the spectral filter in the heralding arm should have an exponentially decaying impulse response with the response time $t_\text{m}$.
Such a filter can be implemented using a high-finesse optical cavity or a combination of optical cavities \cite{Haase2009, Qin2015}. 
The transmission function of the required filter and its impulse response read, respectively:
	\begin{align}
	\begin{split}
	& F(\w) = \left(1-2\ii\w t_\text{m}\right)^{-1}, \\
	& \F(t ) = \frac{1}{2t_\text{m}} \ee^{-t/2t_\text{m}} \Theta(t), 
	\end{split}
	\L{F_filter}
	\end{align}
where $\Theta(t)$ is the Heaviside step function.
Substituting (\ref{psi_spdc}) and (\ref{F_filter}) into (\ref{Phi_ww} - \ref{psi_t-1}) one gets expressions for the joint state of photons after the filter and the heralded state of a signal photon, both in frequency and temporal domains.
Fig.~\ref{fig:shapes}(a) shows the normalized temporal distribution of idler photons calculated as $\int\ud t' |\Psi(t,t')|^2$, using (\ref{Psi_tt}).
Fig.~\ref{fig:shapes}(b) shows the temporal distribution of signal photons heralded upon detection of idler photons at different time instants, which are marked with vertical dashed lines.
The time instants are random and come from the quantum measurement process.
One sees that heralded signal photons are synchronized with the heralding instants.
For heralding instants $t'~\in~(t_\text{m},t_\text{u}]$, the temporal shapes of photons are identical and reproduce smoothed rising exponent, i.e. time-reversed impulse response of the filter placed in the heralding arm.
The smoothing is due to the finite correlation time of the photon pairs.

However, for signal photons heralded in the beginning and the end of the time interval, there are deviations from the properties stated above.
Namely, for heralding instants $t' ~\in~[0,t_\text{m})$, the
photon shape depends on the heralding instant and represents itself a distorted rising exponent.
For heralding instants $t'>t_\text{u}$ the signal photon is not synchronized with the heralding instant.
These deviations are due to the model of entangled photon pairs used (see expression~(\ref{psi_spdc})), namely,  due to the assumption that the photon pairs span only a finite time window.
When the condition $t_\m \ll t_\u$ is fulfilled, then most of the heralded photons have exponentially rising temporal shape.

We now assume that all idler photons, that went trough the filter, are used to herald shaped signal photons.
Then for parameters of the example presented in Fig.~\ref{fig:shapes} the heralding probability reads according to (\ref{RR}): $R=0.17$.
It coincides well with an estimation obtained from (\ref{wm/wu}).
For comparison, the probability to herald shaped signal photons using the temporal modulation and frequency resolving detection of idler  photons, is estimated as $R' \approx 0.02$ (see \cite{Averchenko2017}).
Thus, the first method allows to achieve the heralding probability, which is ten times higher than the heralding probability achieved by the second method, when the same sources of entangled photons are used.

	\begin{figure}
	\center{\includegraphics[width=0.95\linewidth]{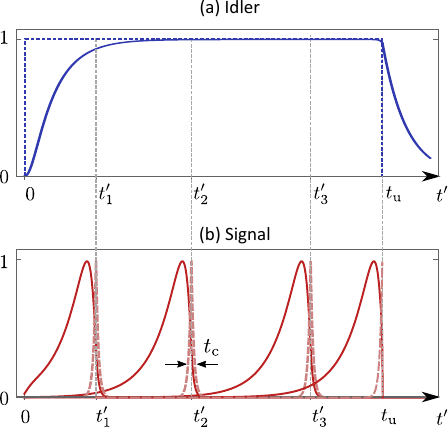}}
	\caption{
	(a) Temporal distribution of idler photons before (dashed line) and after (solid line) spectral filter in the idler arm.
	(b) Temporal distribution of signal photons conditioned upon detection of idler photons at time instants $t_1', t_2', t_3'$ (solid curves).
	Dashed curves represent distributions obtained without spectral filtering of idler photons and reproduce second-order cross-correlation function.
	The functions are normalized to unity at their maximum.
	The relation between parameters is chosen as following: $t_\u:t_\m:t_\c = 150:10:1$.
	The heralding probability for these parameters is $R \approx 0.17$ according to (\ref{RR}).
	}
	\L{fig:shapes}
	\end{figure}

\section{Generation of shaped photons from a continuous stream of entangled photon pairs}\L{cw}

Here we consider a more advanced model to describe the generation of photons with a given temporal profile using nonlocal spectral filtering.
We consider spatially single-mode signal and idler fields with slowly varying amplitudes $\hat a_\text{s,i}(t)$.
Squared modules of these amplitudes define fluxes of signal and idler photons.
We use the Heisenberg formalism to describe the transformation of the fields during the shaping procedure.
We assume that the fields are stationary and their properties are fully characterized by the following first order auto- and cross-correlation functions, respectively (see, for example, \cite{Averchenko2018}):
	\begin{align}
	& G_\text{ss,ii}^{(1)}(t,t') = \<\hat a_\text{s,i}^\+(t) \hat a_\text{s,i}(t')\> = \bar n \ee^{-\frac{|t-t'|}{2 t_\c}} \left(1+\frac{|t-t'|}{2 t_\c}\right), \L{Gss}\\
	& G_\text{si}^{(1)}(t,t') = \<\hat a_\text{s}(t) \hat a_\text{i}(t')\> = \sqrt{\frac{\bar n}{2 t_\c}} \ee^{-\frac{|t-t'|}{2 t_\c}}. \L{Gsi}
	\end{align}
Right hand sides represent correlation functions of fields which are generated via SPDC or SFWM process in a symmetric low-loss nonlinear optical cavity in the low pump regime (see, for example, \cite{Fortsch2013, Lu2016}). 
The expressions imply that the average fluxes of signal and idler photons are $\bar n$ and the correlation time of the fields is $t_\c$ . 

Filtering of the idler field performed in the shaping method is described by the following transformation of the field amplitude similar to (\ref{t_filt})
	\begin{align}
	& \hat a_\i(t) \rightarrow \int  \ud\t \F^*(\t) \; \hat a_\i(t+\t),
	\L{a_filt}
	\end{align}
where $\F(\t)$ is the impulse response of the filter.
The average flux of idler photons after the filter can be obtained by combining (\ref{a_filt}) and (\ref{Gss}).
Particularly, one gets the following explicit expression for the photon flux when a filter with the exponentially decaying impulse response is used (\ref{F_filter}):
	\begin{align}
	& \bar n_\i' = G_\text{ii}^{(1)}(t,t) = \bar n
 \frac{(2t_\c^{-1}+ t_\m^{-1})}{t_\m(t_\c^{-1}+t_\m^{-1})^2} = [t_\m \gg t_\c] \approx	 \bar n \frac{2 t_\c}{t_\m}.
  	\end{align}
The expression shows that the flux of idler photons after the filter is reduced by $\frac{2 t_\c}{t_\m}$ times.
Furthermore, the expression gives the heralding probability of signal photons, if all idler photons after the filter are used as heralds:
	\begin{align}
	R &= \frac{\bar n_\i'}{\bar n} 
	=  \frac{(2t_\c^{-1}+ t_\m^{-1})}{t_\m(t_\c^{-1}+t_\m^{-1})^2} \L{R_CW}\\
	&= [t_\m \gg t_\c] \approx	 \frac{2 t_\c}{t_\m}
	\L{R_CW-1}
 	\end{align}
It confirms the estimation of the heralding probability obtained above (\ref{wm/wu}).

The cross-correlation function of signal and idler fields after the idler filtering is obtained by substituting (\ref{a_filt}) into (\ref{Gsi}).
The coincidence detection of signal and idler photons is characterized by the following normalized second-order cross-correlation function of the signal and idler fields $g_\si^{(2)}(t,t')$:
    \begin{widetext}
	\begin{align}
	g_\si^{(2)}(t,t')  = 1+\frac{|G_\si^{(1)}(t,t')|^2}{n_\s(t)n_\i'(t')}
	=1 + \frac{1}{2\bar n t_\c} \frac{1}{1+\frac{2 t_\m}{t_\c}} 
	\left\{\begin{array}{ll}
	\left(\frac{2-\left(1+\frac{t_\c}{t_\m}\right) \ee^{\left(1-\frac{t_\c}{t_\m}\right)\left(t-t'\right)/2t_\c} }{1-\frac{t_\c}{t_\m}}\right)^2\ee^{(t-t')/t_\m}, & t \leqslant t'\\
	\ee^{-(t-t')/t_\c} & t>t'
	\end{array}\right.
	\L{CW_photon}
	\end{align}
	\end{widetext}
The expression shows that the correlations are stationary after the filtering of the idler field.
The first term of the function describes accidental coincidence detection of photons and the second term describes coincidence detection due to the detection of photons of a correlated photon pair. 
The later reproduces smoothed time-inverted impulse response of the filter, under the condition $t_\c \ll t_\m$.
Furthermore, the second term exceeds the first one when 
	\begin{align}
	& t_\m \ll \bar n^{-1},
	\L{tmn}
	\end{align}
i.e. when the number of photon pairs per response time of the filter is smaller than one. This condition reproduces the right-hand side of the condition (\ref{t-cond}), provided a continuous stream of photon pairs is used.

Summing up, the obtained result confirms the interpretation 
that the detection of idler photon in the considered shaping method heralds a signal photon with the temporal shape determined by the impulse response of the filter. The photon shape does not depend on the heralding instant, but it is synchronized with the heralding instant. When one uses a continuous stream of narrowband photon pairs with the properties described by the functions (\ref{Gss}, \ref{Gsi}) and apply a filter with the exponentially decaying impulse response (\ref{F_filter}), then a non-normalized temporal shape of the heralded photon reads:
	\begin{align}
	& \psi_\s(t|t') \propto  \left\{\begin{array}{ll}
	\left(\frac{2-\left(1+\frac{t_\c}{t_\m}\right) \ee^{\left(1-\frac{t_\c}{t_\m}\right)\left(t-t'\right)/2t_\c} }{1-\frac{t_\c}{t_\m}}\right)\ee^{-(t'-t)/2t_\c}, & t \leqslant t'\\
	\ee^{-(t-t')/2t_\c} & t>t'.
	\L{exp-CW}
	\end{array}\right.
	\end{align}

\section{Excitation of a two-level atom with an exponentially shaped photon}	\L{sec:atom}

Here we consider the excitation of a two-level atom through $4\pi$ solid angle illumination with an exponentially shaped single-photon pulse (\ref{exp-CW}).
It is known that the probability of finding the atom in the excited state depends on the photon temporal shape and is maximal for the exponentially rising shape with the rise time of the shape matched to the atom radiative lifetime \cite{Stobinska2009, Wang2011,Leong2016}. 
Maximum excitation probability can be used to quantify  the quality of the exponential shape of single photons generated via the described above method of nonlocal filtering.

To estimate the excitation probability we use the analogy between the dynamics of the excitation of a two-level atom with a single photon and loading of a photon into a single-sided optical cavity \cite{Heugel2010, Bader2013}. 
Then the excitation probability at the time instant $t$ reads as
	\begin{align}
	& p(t) = \int\limits_{-\infty}^t \left(|\psi_\s(t')|^2 - |\tilde\psi_\s(t')|^2\right) \ud t',
	\L{R}
	\end{align}
where $\psi_\s(t)$ is the temporal shape of the photon and $\tilde\psi_\s(t) = (2/t_\m) \int \ee^{-(t-t')/t_\m} \psi_\s(t') \ud t' - \psi_\s(t)$ is a shape of the photon scattered by the atom.
Fig.~\ref{fig:ptt} shows the time-dependent probability of finding a two-level atom in the upper state which is excited with shaped single photon (\ref{exp-CW}) heralded at the time instant $t'=0$.
The temporal shape of the photon is shown in Fig.~\ref{fig:ptt} by the dashed line.
The values of $t_\c$ and $t_\m$ are the same as in Fig.~\ref{fig:shapes}. 
Notably, the moment of maximum excitation probability is synchronized with the heralding instant of the shaped photon (in Fig.~\ref{fig:ptt} this moment is $t=0$).

The maximum excitation probability $p_\text{max} = \text{max}_{t} p(t)$ of the atom with the shaped photon (\ref{exp-CW}) can be calculated analytically
	\begin{align}
	& p_\text{max} = \frac{\e}{\e+1/2},
	\quad \text{where}\quad \e = t_\m/t_\c.
	\L{epsilon}
	\end{align}
	\begin{figure}
	\center{\includegraphics[width=0.9\linewidth]{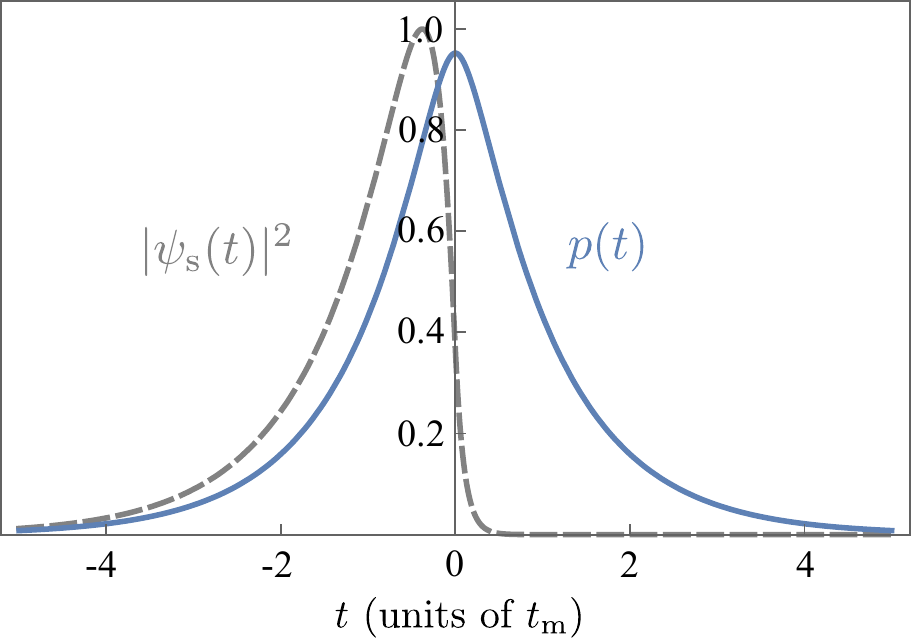}}
	\caption{
Time-dependent probability $p(t)$ (solid blue line) of finding a two-level atom in the upper state upon free space mode-matched excitation of the atom with a single-photon resonant pulse with exponentially rising temporal shape $|\psi_\s(t)|^2$ (gray dashed line) produced via nonlocal filtering (see Exp.~(\ref{exp-CW})).
The parameters are chosen the same as in Fig.~\ref{fig:shapes}, i.e. $t_\m:t_\c = 10:1$. Temporal shape of the photon is normalized to unity at the maximum.  
Maximum excitation probability for given parameters is $p_\text{max} = 0.95$ and it is achieved at $t=0$ that is the heralding instant of the shaped photon.
	}
	\L{fig:ptt}
	\end{figure}
Fig.~\ref{fig:r} shows both the maximum excitation probability and the heralding probability of the shaped photons (\ref{R_CW}) as functions of the parameter $\e$.

	\begin{figure}
	\center{\includegraphics[width=0.9\linewidth]{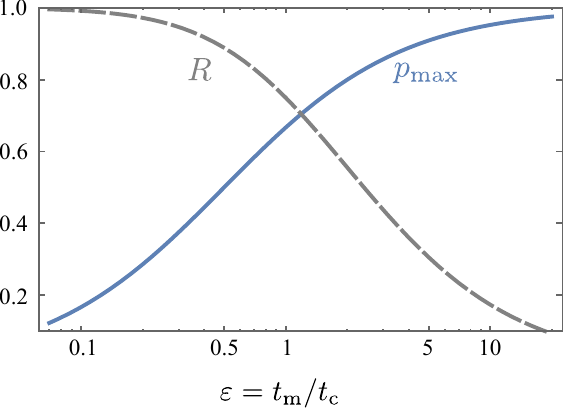}}
	\caption{
	Gray dashed line denote heralding probability (\ref{R_CW}) of photons with the exponentially rising temporal shape (\ref{CW_photon}) as a function of $\e = t_\m/t_\c$. 
	Solid blue line denotes maximal excitation probability (\ref{epsilon}) of a two-level atom achieved via the full solid angle illumination with the single photon. The rise time of the exponential photon shape is matched to the radiative lifetime of the atomic transition. 
}
	\L{fig:r}
	\end{figure}

\section{Influence of potential imperfections}\L{sec:imperfections}

Here we consider several experimental imperfections which can affect the properties of the shaped photons heralded using nonlocal spectral filtering and we present experimental conditions to reduce effects of these imperfections.

First, the finite temporal resolution of the detector $t_\text{d}$ in the heralding arm (see Fig.~\ref{fig:scheme}(b)) results in a corresponding uncertainty of the heralding instant of the  shaped signal photon.
The effect is negligible if the resolution $t_\text{d}$ is shorter than the temporal duration of the heralded photon defined by the temporal response of the filter, i.e. $t_\text{d} \ll t_\m$.

Second, a drift of the filter central transmission frequency by $\w_\text{d}$ affects the spectral and temporal properties of the heralded photon. 
Particularly, in the limit of perfect frequency anti-correlations of the photons, a positive shift of the filter  frequency by $\w_\text{d}$ results in a negative shift of the spectrum of the heralded photon by the same amount as can be seen from Fig.~(\ref{fig:shape}).
The effect is negligible if the frequency shift is smaller than the bandwidth of the heralded photon, i.e. $\w_\text{d} \ll t_\m^{-1}$.

\section{Estimations of achievable parameters}\L{sec:estimations}

Here we consider temporal durations and temporal shapes of photons which are achievable with the discussed shaping method.
Conditions (\ref{t-cond}) and (\ref{tmn}) can be summed up as the following inequalities for the temporal duration $t_\m$ of shaped photons
	\begin{align}
	& \{t_\c, t_\text{d}\} \ll t_\m \ll \{t_\text{coh}, \bar n^{-1}\}
	\end{align}
The duration is limited from below by the correlation time $t_\c$ of photon pairs, which are used as an entangled resource, and by the temporal resolution of the detector $t_\text{d}$ in the heralding arm.
For SPDC-/SFWM-based sources of entangled photons the time is defined by the phase-matching bandwidth in a nonlinear medium (fs-ns range) or by the photon lifetime in a cavity-assisted configuration (ns range). The temporal resolution of modern single-photon detectors is tens to hundreds of picoseconds.
Furthermore, for SPDC-/SFWM-based sources of entangled photon pairs the coherence time $t_\text{coh}$ of the pump laser ($\mu$s-ms range) limits the maximal $t_\m$.
Therefore, discussed shaping method is suitable to produce shaped photons from ns to $\mu$s.
Also the number of photon pairs per time constant of the filter $t_\m$ should be smaller than one, i.e. $\bar n \; t_\m \ll 1$. This limits the flux of shaped photons to MHz$\div$GHz range.

Notably, the impulse response of any passive spectral filter fulfills a general causality condition, namely, the response depends on past and current inputs, but not on future ones. 
This can be formally expressed as $\F(\t<0) = 0$ in (\ref{shape_t}). This condition implies that the shape of the heralded photon always terminates at the heralding instant. In other words, only the shapes with a distinctive end can be produced with the proposed shaping method.

The range of achievable photon durations and a relatively simple way to produce photons with an exponentially rising temporal shape show that the method is particularly useful to produce photons suitable for efficient interaction with single quantum emitters, such as single atoms or ions.
To increase the heralding probability of shaped photons, it is advantageous to use entangled photons, such that their unconditional bandwidth $\w_\u$ is close to the desired bandwidth $\w_\m$ of shaped photons, as it follows from the expression (\ref{wm/wu}).
Thus, such sources of narrowband entangled photon pairs, as cavity-assisted SPDC/SFWM (see, for example, \cite{Schunk2015, Lu2016}) and SFWM in an atomic ensemble \cite{Gulati2014, Liu2014}, are particularly attractive for the efficient generation of shaped photons suitable to excite single quantum emitters.

As an example, we estimate parameters of a source of entangled photon pairs, which is suitable to produce shaped photons for the efficient photon-atom coupling. We set the maximum excitation probability to be $p_\text{max} \approx 90\%$ and the pulse duration to be $t_\m = 35$ns.
It follows from (\ref{epsilon}) and Fig.~\ref{fig:r} that photon pairs with a correlation time of not more than $t_\c = 7$ ns are required.
Furthermore, the flux of entangled photon pairs is limited by the condition $\bar n \ll t_\m^{-1} = 30$~MHz according to (\ref{t-cond}).
The heralding probability for the above parameters is $R \approx 0.3$ (see Fig.~\ref{fig:r}) and the average flux of heralded shaped photons is limited by $\bar n_\s \ll \bar n R = 10$ MHz.

\section{Conclusions}\L{sec:conclusion}

We studied a method to produce temporally shaped photons from entangled photon pairs via nonlocal spectral filtering. 
Namely, a temporally shaped photon is produced via spectral filtering and time resolving detection of its entangled counterpart.
We establish conditions on the entanglement properties, parameters of the spectral filter and temporal resolution of the detector, which ensure that the shape of the heralded photon is defined by the time-reversed impulse response of the spectral filter in the heralding arm.

We show that the method has several distinguishing features as compared to shaping via the nonlocal temporal modulation \cite{Averchenko2017}.
First, heralded photons are synchronized with the heralding instants and have the same temporal shape.
Second, post-selection of outcomes of a time-resolving single-photon detector is not required to produce identically shaped photons with the same central frequency.
As a result, the shaped photons are produced at a much higher rate, exceeding the one attainable by nonlocal temporal modulation of photons \cite{Averchenko2017}.
Third, when a heralded shaped photon is used to excite a two-level atom in free space then the moment of the maximum excitation probability is synchronized with the heralding instant. 
Fourth, the method imposes very moderate requirements on the experimental resources. Given a source of entangled photons, the method requires just two shaping components to produce shaped photons: a spectral filter and a time-resolving photon detector. 
Finally, the method provides shaped photons in the broad range of durations from ns to $\mu$s. 
This method is particularly suitable for producing photons with a rising exponential shape, which are required for efficient light-matter quantum interface.
For example, one can use an optical cavity as a shaping filter, which has required exponentially decaying impulse response.

The studied heralded shaping via the spectral filtering
alongside with the shaping method based on the temporal modulation
\cite{Averchenko2017} constitute a toolbox to produce shaped
photons with different durations and shapes in a heralded and potentially scalable way.
We anticipate that such methods will push forward quantum optics
experiments with temporally shaped photons.

We thank Paul Kwiat, Markus Sondermann, Gerhard Schunk, Ulrich Vogl for useful discussions. 
VA acknowledges funding by Max-Planck-Gesellschaft and Russian Science Foundation grant No. 17-19-01097. 
DS acknowledges Russian Science Foundation grant No. 17-72-30036 (Sec.~2, analysis of quantum measurements).

\bibliography{w_shaping}

\end{document}